\documentclass[twocolumn]{article}
\usepackage[authoryear,round,longnamesfirst]{natbib}

\usepackage{amssymb}
\usepackage{graphicx}
\usepackage{caption}
\usepackage{sectsty}
\usepackage{amsmath}
\usepackage{url}
\sectionfont{\fontsize{10.5}{10}\selectfont}

\providecommand{\keywords}[1]{\textbf{\textit{keywords---}} #1}
\newcommand{\twobar}{/\kern-0.2em/}
\title{The Multiperiodic Pulsating Star {\it Y~Cam~A} as a Musical Instrument}

\author{B. Ula\c{s}\\ {\small \.{I}zmir Turk College Planetarium, 8019/21 sok., No: 22, \.{I}zmir, Turkey}\\ {\small bulash@gmail.com}}

\begin{document}

\maketitle

\begin{abstract}
In this study we generate musical chords from the oscillation frequencies of the primary component of oscillating eclipsing Algol system Y~Cam. The parameters and the procedure of the musical chord generation process from the stellar oscillations are described in detail. A musical piece is also composed in appropriate scale for Y~Cam~A by using the generated chords from the results of the asteroseismic analysis of the stellar data. The music scores and the digital sound files are provided for both the generated chords and the musical composition. Our study shows that the further orchestral compositions can be made from the frequency analysis results of several pulsating stars by using the procedure stated in present study.
\end{abstract}


\section{Introduction}

The efforts on combining the celestial objects with music have been made for ages. Pythagorians\textsc{\char13} philosophical concept, Musica Universalis, pointing a harmony in celestial movements and Kepler's works on relating the planetary motions to musical consonance can be given as examples. There are also modern studies focused on combination of the music with pulsating stars in more mathematical and efficient ways. For instance, in their extensive work, \cite{kol06} made a musical composition by developing a software to transform the stellar oscillations that can be based on the note $C$. The authors considered the internal structures of stars while determining their criteria in selecting some certain pulsating stars as musical instruments. They also analyzed their composition in musical point of view. An agreement between musical scales and the frequencies of multiperiodic stellar pulsations were investigated by \cite{ula09}. The author concluded that the best agreement was reached for the binary system Y~Cam and the Diminished Whole Tone Scale. Various digital sound files made by transforming the stellar pulsations to audible range also can be reached online\footnote{\ttfamily{https:\twobar goo.gl/FFhpEV}}.

We describe the chord generation from the results of the frequency analysis of a pulsating star in detail in the next section. The third section focuses on a musical composition in an appropriate musical scale made by using a digital piano and the generated chords from the oscillations of the primary component of Y~Cam. We concluded the results in the last section.

\section{Chord Generation from Stellar Oscillations}

A chord is defined as simultaneous sound of three or more different tones which are formed by following the certain intervals \citep{ape44}. It can be thought that a chord is a group of sound waves having many frequencies emerged simultaneously from a musical instrument. This phenomenon shows analogy to the light received from a multiperiodic pulsating star. The light curve of a multiperiodic pulsating star exhibits a wave containing many sine like variations with certain frequencies, amplitudes and phase shift values which can be determined by analyzing the light emerged from the star. This similarity were used in generating a chord from the asteroseismic results of the multiperiodic pulsating component of Y Cam in this study. 

We first defined three dimensionless transformation parameters in order to generate a musical chord from the stellar oscillations: The relative frequency was defined as $f^{\prime}=\frac{f_{i}}  {f_{min}}$ where $f_i$s are the derived oscillation frequencies for a pulsating star and $f_{min}$ is the minimum frequency value among the derived ones. The loudness parameter,  $L=\frac{A_{i}}  {A_{max}}$, refers to the relative loudness of a signal and it is the ratio of $A_{i}$, amplitude value for a given oscillation frequency of star, to $A_{max}$, maximum amplitude value within the derived frequency group. $p$ parameter denotes the starting time of the signal and is the difference between the phase shift value of a given frequency and the minimum phase shift value $(p=\phi_{i} - \phi_{min})$ among the derived parameters from oscillations of the star. 

We calculated the three parameters $(f^{\prime}, L, p)$ for the oscillation frequencies of Y~Cam~A derived by \cite{kim02}. Following their results we calculated the necessary quantities for the transformation as $f_{min}=15.0473~c/d$, $A_{max}=5.8~mmag$ and $\phi_{min}=-1.31$. In order to generate a chord which is unique for the star we first generated the tones for the mentioned chord using the necessary parameters described above. Table~\ref{tabchord} lists the calculated values for the parameters which were used to generate the tones by using the {\sc Audacity}\footnote{\ttfamily{http:\twobar www.audacityteam.org/}} audio editor. The program can generate tones for a given frequency, normalized amplitude and starting time values. Three steps in the process of tone generation were followed: {\it (i)} We multiplied all $f^{\prime}$ values with the frequency value of the desired musical note. For instance, when we wanted to transform the smallest frequency to the fourth octave note $G$ we multiplied $f^{\prime}$ values with 392.0~Hz\footnote{\ttfamily{http:\twobar www.phy.mtu.edu/$\sim$suits/notefreqs.html}} and we derived four new frequency values (392.00~Hz, 398.00~Hz, 469.09~Hz and 483.65~Hz). These values were applied to the program as the frequencies of four tones. {\it (ii)} The $L$ values were directly entered the program as normalized amplitudes. {\it (iii)} the $p$ parameters were the input values characterizing the starting times of the generated tones. When finished with those three steps we generated four tones (one for each oscillation frequency) which can be played together to generate the chord on the desired note. The same procedure was followed for the notes $G$, $A$, $C$ and $D$. The frequency values of the tones based on these notes are listed in Table~\ref{tabchord}. A group of digital sound files for the generated chords on the mentioned notes were also produced and can be found in an online playlist\footnote{\ttfamily{https:\twobar soundcloud.com/bulash/sets/chordsycamb}}

\begin{table*}
\centering
\caption{The oscillation properties of Y~Cam together with the calculated transformation parameters of chord generation. The second, third and fourth columns list the frequency, amplitude and phase shift values of the pulsating component derived by \cite{kim02}. Headers of the last four columns indicate the musical notes whose frequencies were multiplied by $f^{\prime}$ values when generating the tones for the chord. The indexes $4$ and $5$ refer the octave number of the notes, i.e. $C_5$ means the fifth octave note $C$. An octave is the interval between first and the eighth tones of eight consecutive diatonic tones \citep{bak04}.}
\label{tabchord}
\begin{tabular}{lcccccccccc}
\hline
	&	$f~(c/d)$	&	A(mmag)	&	$\phi$	&	$f^{\prime}$	&	$L$	&	$p$	&	$G_4 (Hz)$	&	$A_4 (Hz)$	&	$D_5 (Hz)$	&	$C_5 (Hz)$	\\
\hline
$f_1$	&	15.0473	&	5.8	&	-0.89	&	1.0153	&	1.0000	&	0.4200	&	398.00	&	446.74	&	596.33	&	531.26	\\
$f_2$	&	18.2852	&	3.9	&	1.06	&	1.2338	&	0.6724	&	2.3700	&	483.65	&	542.87	&	724.64	&	645.58	\\
$f_3$	&	14.8203	&	3.4	&	-1.31	&	1.0000	&	0.5862	&	0.0000	&	392.00	&	440.00	&	587.33	&	523.25	\\
$f_4$	&	17.7348	&	2.8	&	-0.35	&	1.1967	&	0.4828	&	0.9600	&	469.09	&	526.53	&	702.83	&	626.15	\\
\hline
\end{tabular}
\end{table*}

\section{Musical Composition using Stellar Chords}

\begin{figure}
\centering
\includegraphics[scale=1.2]{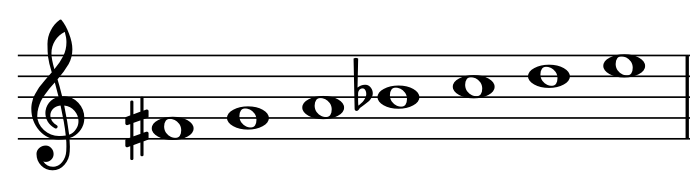}
\caption{The Diminished Whole Tone Scale on $F\sharp$ in staff with the $G$ clef.}
\label{scale}
\end{figure}

Since the oscillation frequencies of the primary component of the eclipsing binary system Y~Cam shows an agreement with the Diminished Whole Tone Scale \citep{ula09} we planned to compose a piano piece in mentioned scale and accompany the chords generated from the oscillations of the star in the previous section. A musical scale is defined as series of countless variety of selected musical notes \citep{woo64}. The Diminished Whole Tone Scale on $F\sharp$ was selected as the scale during the composition. The places of the musical notes in Diminished Whole Tone Scale on $F\sharp$ on the staff is given in Fig.~\ref{scale}. The {\sc MuseScore}\footnote{\ttfamily{https:\twobar www.musescore.org/}} software is used to form a sheet music for the composed piece (Fig.~\ref{sheet1} and ~\ref{sheet2}). The program allows the user to put numerous musical notations on a blank sheet in digital media to prepare a sheet music. We created a digital sound file for the composition in two steps. First a sound file\footnote{\ttfamily{https:\twobar soundcloud.com/bulash/akycampiano}} was created by recording the piano performing by the author; then, the recorded sound file mixed with the previously generated stellar chord files following the notation on the sheet music for the composition. The mixing process was carried out using the digital audio editor {\sc GoldWave}\footnote{\ttfamily{http:\twobar www.goldwave.com/}}. The resulting sound file for the final composition can be found online\footnote{\ttfamily{https:\twobar soundcloud.com/bulash/akycam}}.

Fig.~\ref{sheet1} and \ref{sheet2} demonstrates the music scores of the whole musical composition. First two staffs from the top (accolade) on the sheet show the piano partition of the piece. The third staff with unusual notation contains the star signs locate on the staff lines and refer to the notes for the smallest frequency of the generated chord, namely the smallest oscillation frequency of Y~Cam~A after it was transformed to the desired musical notes following the directions described in the previous section. The star signs on the third staff also show the places where the chords are played through the composition. The duration of the chord playing was set according to selected time of the composition, three-four time. This type of notation for the third staff was used because of the impossibility of displaying the second, third and fourth frequencies of the generated chords by using any musical notation on staff with the $G$ clef.

\begin{center}
\begin{figure*}
\ContinuedFloat*
\includegraphics[scale=0.6]{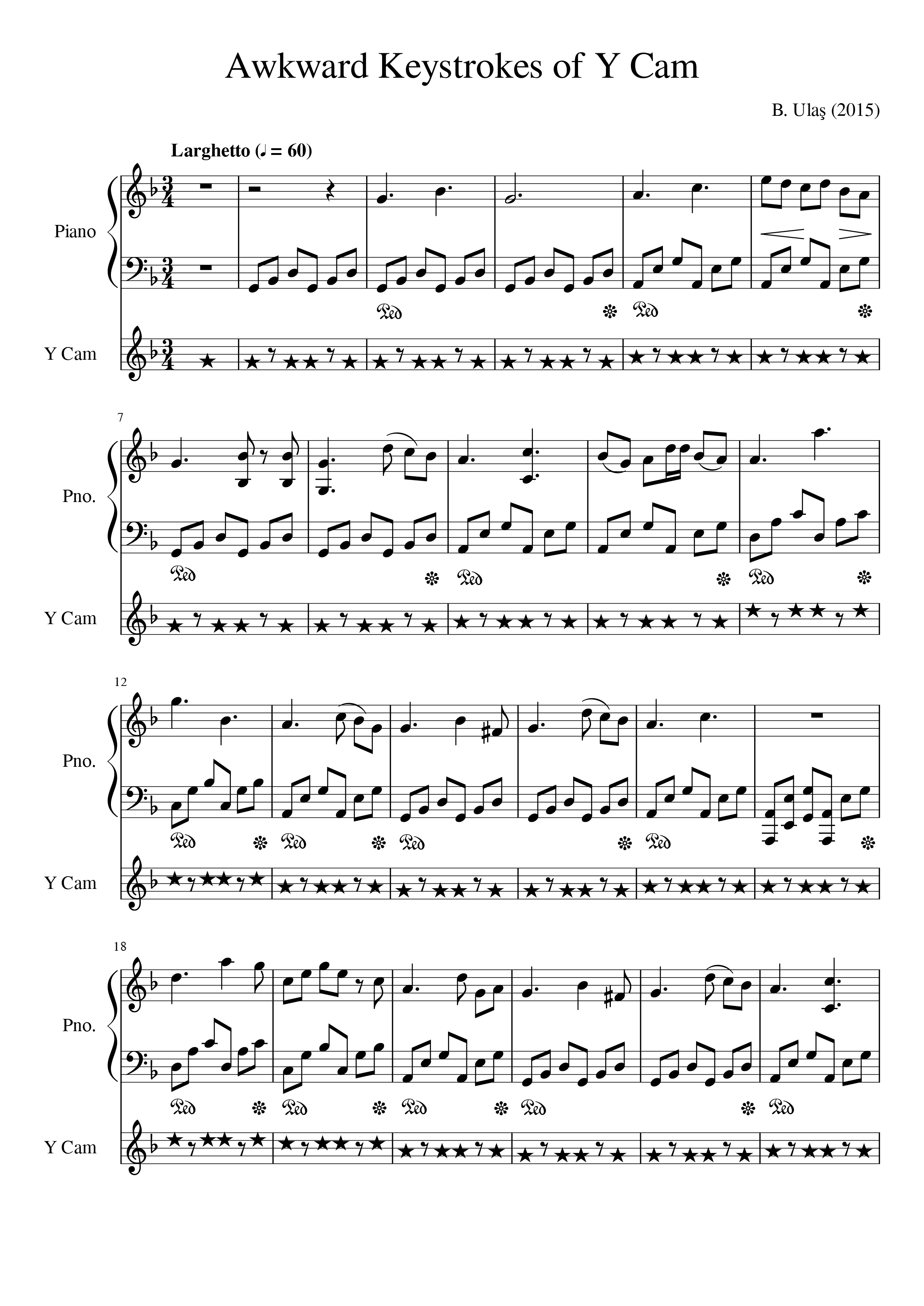}
\caption{Sheet music for the composition.}
\label{sheet1}
\end{figure*}
\begin{figure*}
\ContinuedFloat
\includegraphics[scale=0.6]{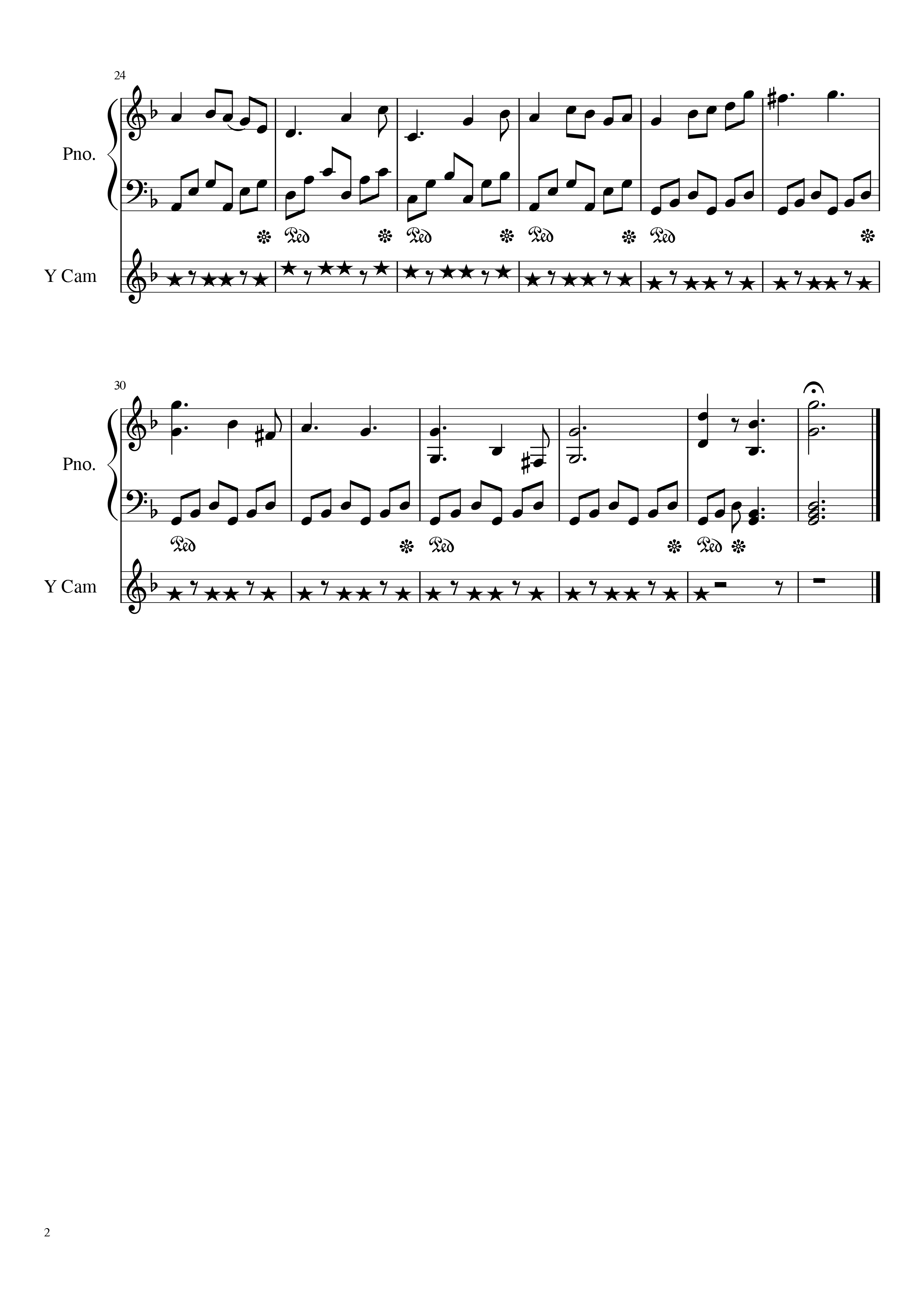}
\caption{(Cont.)Sheet music for the composition. }
\label{sheet2}
\end{figure*}
\end{center}

\section{Conclusion}

A short musical composition was made by using the frequency analysis data of the binary system Y~Cam \citep{kim02} whose oscillation frequency arrangement of the primary component shows the best agreement with the Diminished Whole Tone Scale according to \cite{ula09}. We also explained the general properties of the transformation procedure for the oscillation frequencies of pulsating stars to the musical chords on desired note. The procedure is based on the mathematical analogy between musical chords and the observed frequency group of a multiperiodic pulsating star.

The whole procedure can be summarized as converting the oscillation phenomenon seen on the primary component of Y~Cam to a musical instrument which plays its unique chord on various notes. These chords were placed proper to the tempo and the time of the composition which is composed in appropriate scale and given in Fig.~\ref{sheet1} and \ref{sheet2}. The quality of the composed music is controversial, however, the basic steps explained here for one piano and one pulsating star can also be applied to make an orchestral composition using several musical instruments and pulsating stars which will enrich the quality and the listening for the audiences. The digital artificial effects  can also be added to orchestral composition by using contemporary sound editors. The present study was planned to be grown in that route.


\end{document}